# Internet-Based Remote Collaboration at the Neutron Residual Stress Facility at HFIR

M.C. Wright, C.R. Hubbard, R. Lenarduzzi and J. A. Rome
Oak Ridge National Laboratory

## Abstract

The Neutron Residual Stress Facility (NRSF) at ORNL's High Flux Isotope Reactor (HFIR) is being significantly upgraded in conjunction with the upgrade of the HFIR and associated neutron scattering facilities. We have rewritten the LabVIEW-based data acquisition and control software for the NRSF to provide much greater flexibility in operation of the new equipment and to allow remote control over the Internet. The software also now controls a large-sample X-ray diffractometer. The user interface is dynamically adapted to the specifics of each instrument based on instrument configuration information contained in external configuration files.

Internet-based remote control is available via three methods. First, a server at the instrument provides local control and accepts connections via an IP socket. Second, instrument status is available on the Web using LabVIEW's built-in web server that will transmit JPEG images of instrument front panels. Third, complete remote control of the data acquisition system is also available to the instrument scientist via Timbuktu Pro, a commercial remote control program.

We also support remote scientific collaboration, not just remote instrument control. We have participated in the development of multiple Electronic Notebooks, implemented as a collection of web pages that are created and viewed through a web browser. Access to the notebooks is controlled via X.509 certificates. We operate a Certificate Authority and issue our own certificates. All access control and session management is performed on the server via Java servlets. This approach requires no cookies on the client's machine or URL-rewriting, and all transmissions are encrypted.

## Introduction

The Neutron Residual Stress Facility[1] (NRSF) at the High Flux Isotope Reactor[2] (HFIR) at Oak Ridge National Laboratory (ORNL) is used for mapping of residual strain gradients within research and engineering size specimens (dimensions: mm to meter) due to joining, thermal processing, deformation processing, internal interfaces, etc. The facility previously shared time on a triple axis neutron spectrometer but two completely new, interchangeable goniometer systems have been built and will be installed this winter in conjunction with the two-year long upgrade of the reactor and its associated neutron scattering instrumentation.

The previous data acquisition software ran on LabVIEW[3] on a PowerMac with motors controlled by several nuBus cards housed in an expansion chassis. Seven position sensitive detectors were readout through a GPIB interface. The motion control electronics for the new systems are interfaced through GPIB allowing us to move to a Windows platform. We have rewritten the data acquisition and control software to provide much greater flexibility in operation of the two new instruments and to allow improved remote control over the Internet. The rewritten software also controls a new large-sample X-ray diffractometer used for residual stress mapping on surfaces. The user interface is dynamically adapted to the specifics of each instrument based on instrument configuration information contained in external configuration files.

Development of Internet-based remote operation of the NRSF was begun in earnest as a part of the Materials Microcharacterization Collaboratory[4] (MMC), a pilot project funded by US Department of Energy's DOE2000[5] program. The DOE2000 program had three main goals: improved ability to solve DOE's complex scientific problems, increased R&D productivity and efficiency, and enhanced access to DOE resources by R&D partners. One of the strategies to meet these goals was the construction of national collaboratories. These provide the integration on the Internet of unique or expensive DOE research facilities and expertise for remote collaboration, experimentation, production, software development, modeling, or measurement. In addition, collaboratories benefit researchers by providing tools for video conferencing, shared data-viewing, and collaborative analysis. While the MMC was centered around electron microscopy, beamline facilities also played an important role in the project.

## Configuration For Control Of Multiple Instruments

The first version of the NRSF control system, "Run the System", was written in 1992 in LabVIEW 2 running on a Macintosh II. Global variables and programmatic control of object properties were not available. Program size and execution speed were significant considerations. As a result, the configuration of the neutron diffractometer was hard-coded into the program. The new version of Run the System uses configuration data files to specify the parameters of the motion axes, including name, units, limits, speed, and hardware type. These parameters are stored in a global array for use throughout the program. Property nodes are used modify the relevant controls and indicators, such as the strings that label the ring control the experimenter uses to select the axis he would like to move. Addition or removal of motion hardware requires modification only to the configuration data file and not the LabVIEW code.

As part of the initialization process, Run the System presents the experimenter with a front panel containing several controls to specify the configuration of the instrument. One control selects the name of the configuration data file. A ring control is used to select which instrument is being controlled. Options currently are: the X-Y-Z neutron diffractometer, the kappa goniometer neutron diffractometer, the large-specimen X-ray diffractometer, and the instrument simulator. We will add an offline alignment system in the future. The simulator is used for debugging the user interface and program flow without the need to be connected to an instrument. The program's top-level screen is modified to display only the operations that are relevant to the selected instrument. The experimenter can also select the type of the multichannel analyzer and the number of detectors to use. Some use is made of LabVIEW's ability to dynamically load sub-VI's for different configurations, but computers are now fast enough and have enough memory that it is not necessary to do that in this application.

## Three Approaches to Remote Control

Tools for interactive instrument control fall into three broad categories: Web-based, remote computer control, and client-server. Each of the three approaches has advantages (and disadvantages) for different applications. Web-based tools show great promise for wide distribution of basic instrument operation, but not all instrument functionality can be easily provided in the confines of a browser. Commercial remote control software can easily control an existing instrument's computer rather than the instrument itself, but remote control software uses proprietary protocols, and there are concerns about security and network efficiency. Client-server distributed computing is efficient and scaleable but requires a complete redesign of an instrument's data acquisition system if it is applied to an existing instrument.

**Web-based Remote Operation**

The World Wide Web is experiencing explosive growth in both its distribution and its capabilities. While originally used for the display of multimedia information, the Web has now become a mechanism for interactive applications. The advantages of a web-based solution are numerous. Web communication is (nearly) platform independent, browsers are ubiquitous and everyone knows how to use them, web traffic is usually allowed to pass through firewalls, and strong security can be implemented. The requirements for remote operation of beamline facilities have many aspects in common with other types of facilities. We desire platform independent interfaces for motion axes and instrument accessories, web or Internet based control, equipment collision protection and safeguards, video image feedback on system condition, and security for instruments and data. Likewise, the requirements for remote collaboration among experimenters at beamline facilities are common. We desire shared data processing and analysis, video conferencing, security for proprietary information and the ability to share use of licensed databases and software.

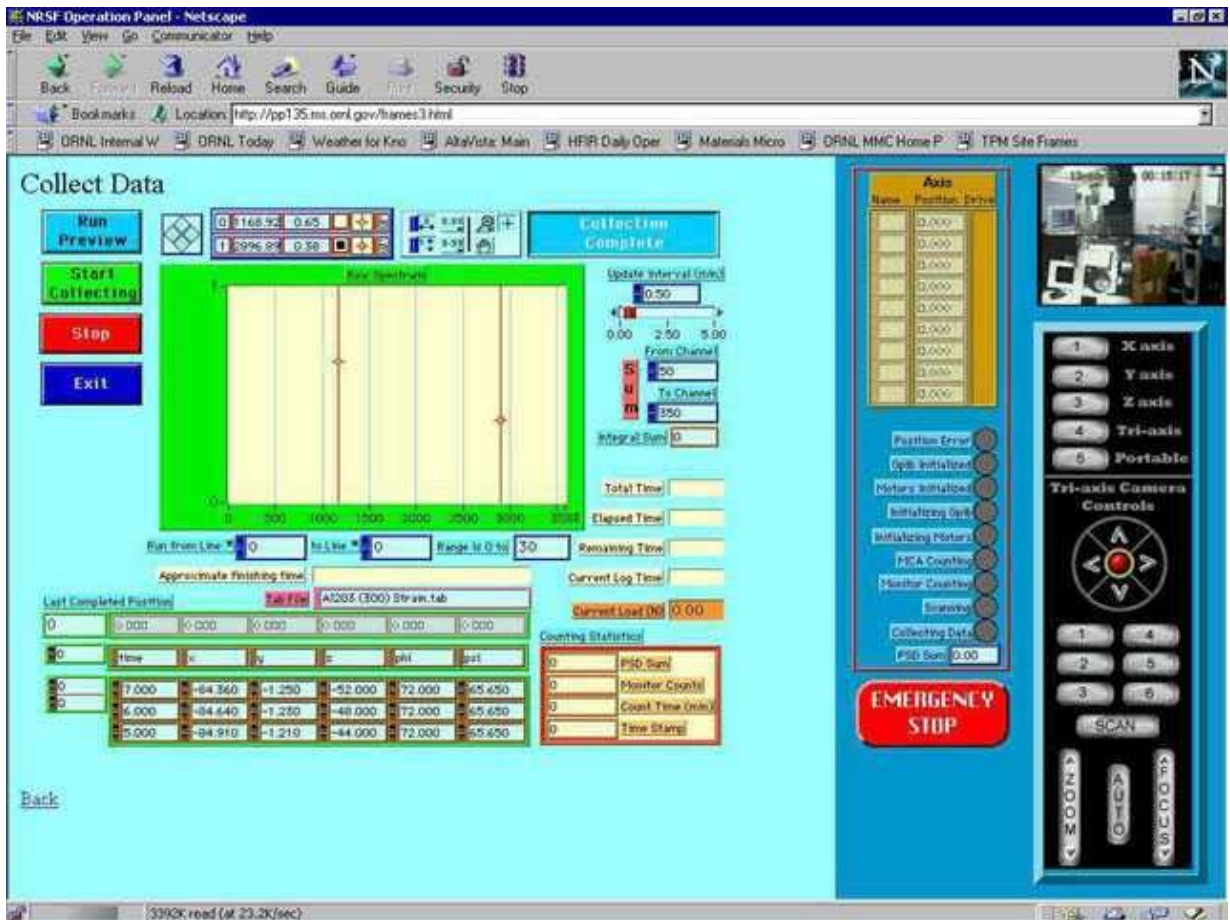

Figure 1 Screen shot of Web-based monitoring of Neutron Residual Stress Facility.

Recent versions of LabVIEW allow the front panel of any open LabVIEW VI (subroutine) to be converted to a JPEG image that is visible through a web browser. The screen shot in Figure 1 shows an example of the remote status screen for NRSF available through the web. The frame on the left is "Collect Data", one of eight VI's that the users frequently access. In the middle is the "Global Status" VI that shows current motor positions and system status. The web server automatically updates these images at a selectable interval, 10 seconds being appropriate for this instrument. These frames show the status of the data acquisition system but do not provide control. For navigation and selection of display of different VI panels, we overlay the images with an image map that allows the user to appear to push the buttons that call and exit the various VI's in the overall system.

The frame on the right in Figure 1shows live video images of the instrument and a control panel to select the camera to use and to control the steerable camera. There are five cameras connected to an RS-232 controlled video switch box that can also combine four of the cameras into the quad image view shown in the figure. The steerable camera is also connected to the computer via RS-232. The steerable camera is important for the remote user to have a sense of "being there" by being able to choose where in the room to look. Three of the fixed camera are positioned on orthogonal axes pointing at the sample. They are equipped with electronically generated cross hairs to precisely indicate the location of the sampling region. The fifth camera is mounted on a flexible goose neck so that it can be easily repositioned to look at a region of interest for a particular experiment.

**Remote Computer Control**

Essentially all modern scientific instrumentation is controlled by a computer and increasingly, these computers are commodity PCs running a version of Windows. In many cases, the controlling software for the instrument is "closed" with no documented ability to externally script or to control the application. Remote control software bypasses this limitation by controlling the remote computer rather than the remote instrument. The screen, keyboard, and mouse of the local computer are duplicated at the remote site. This software is commercially available at low cost. Most of our work has been done using Timbuktu Pro[6] because it supports all Windows versions as well as Macintoshes. A computer running any of these operating systems can control a computer running that same operating system or any of the others. We have productively used Timbuktu Pro for several years for remote control of the NRSF.

Performance on high-speed (>10 Mb/s) local networks is excellent with only a slight sluggishness in the perceived responsiveness of the computer. Performance on a wide area network will, of course, depend on the network performance but is useable even at ISDN speeds. One reason that performance is better than might be expected is that Timbuktu Pro does not simply send bitmap images of the local screen; when possible, it sends the system commands that cause elements of the screen to be redrawn. These commands can require much less bandwidth to transmit than the resulting bitmaps would require.

One key aspect of Timbuktu Pro is that it provides access to the complete remote computer system not just the instrument. This is an advantage for the system owner who can use this feature to restart programs, copy files and perform other functions on the remote computer. However, this same feature is less desirable for remote users who are intended to have only control of the instrument but not have access to the computer's other software and network privileges.

Another limitation of Timbuktu Pro, for some of our purposes, is that it is proprietary and does not support Unix. We would like, for example, to be able to use alternative compression schemes for screen images or to be able to encrypt the communication between the clients and servers. Hooks for these changes are available with VNC[7] from AT&T Laboratories Cambridge. VNC is distributed under the terms of the GNU General Public License, meaning the program and source code are freely distributable. Servers exist for X Windows, Win 32, and Macintosh (beta). A Java viewer is available, which will run in any Java-capable browser. VNC can be used with SSH to provide substantially increased security. VNC is slower than the PC-only products because it does not take advantage of the system screen-drawing commands mentioned above. We are using VNC for remote control of the X14A beamline at NSLS.

### Client-Server Remote Operation

When writing a new data acquisition system, or when sufficient access is provided to an existing one, a client-server approach has many advantages. It can be a very efficient user of network bandwidth and strong security can be readily implemented. We have rewritten our LabVIEW-based data acquisition system to allow a client-server mode of operation. Communication is over an IP socket. Recent versions of LabVIEW provide the ability for LabVIEW applications running on different computers to interact with each other with very little programming required. However, we did not make use of this capability and instead did everything "by hand". One reason is that we want to be able to use clients based on languages other than LabVIEW. For example, we have begun writing a web-based client that communicates over the Internet with our LabVIEW-based server to dynamically provide real-time status information about the system.

## NRSF Data Acquisition System

### Basic Operation

In general terms, both the server and the client application exchange information and commands over the network. The server application directly controls the instrumentation and collects data. It accepts commands from either a local operator or a client that it has granted permission to perform control. The server also broadcasts to all connected clients the status of the application, the values of changed variables, and the diffraction spectrum. The client application duplicates the screen information of the server and updates its display with data received from the server. The client application relies on its own routines to display images identical to the server. If allowed to control, the client can send most of the commands available to the local operator at the server; otherwise it operates in observer mode. The Server must be started first and must have reached its steady state mode, i.e.: must have passed the initialization phase. Then a client can start its version of the application and communicates with the server-based application. The Server keeps a log of the connections and control activity received from the clients.

### Server Application

In addition to the instrument control and data collecting operation, the server performs additional concurrent functions:
- Network connection management,
- Command receiving and processing,
- Broadcasts changes,
- Client connection management on local operator demand.

Multiple clients can connect to the server at the same time. Before the connection is allowed, the client address is compared to a list of approved sites. Once connected, the client receives initializing information (application status and variables values) and operates in OBSERVER mode. Only one of the connected clients can control the operation of the server.

The client can request control privileges. When the server receives the request, it first checks that the requesting client is on the list of clients approved for control, and that control has not already been granted to any other client. If the requesting client is not on the list or control has already been granted, the requesting client is notified accordingly. In this function the server can operate in two modes: manual and automatic check. In automatic mode, if on the approved list, the server grants permission by sending a message to the requesting client. In manual mode, the server prompts the local operator that a client has requested control privileges. The local operator can grant or deny the privilege. The client is then informed of the decision.

Under the network connection management function, the server listens to OBSERVER clients to detect when they have closed their application and therefore disconnected. The server updates the lists of connections accordingly.

The local operator at the server application can select the Manage Clients function, one of the button choices on the screen. This portion of the application allows the operator to:
- View connected clients, their status and which one, if any, has control privileges.
- Remove control privileges temporarily,
- Restore control privileges,
- Remove control privileges permanently,
- Disconnect a client,
- Edit approved site lists.

In Figure 2, the table shows only the clients connected. The controlling client, if any, will be above the line, the observers will be below the line.

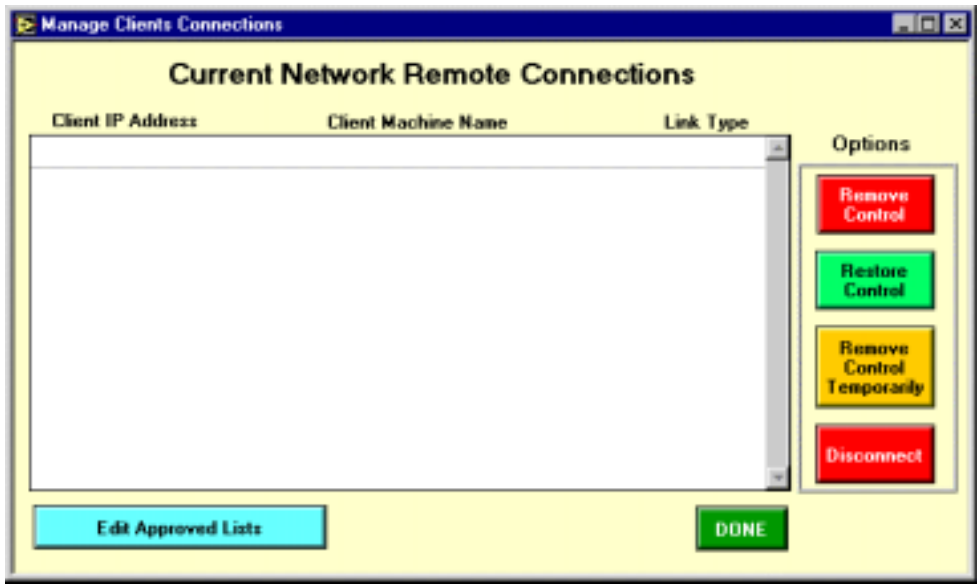

Figure 2 Remote connection management front panel.

Figure 3 shows the edit list function where the operator can add/remove sites to/from the approved list and the control privileges approved list. The lists are stored in a file on the server computer. We will add X.509 certificate-based authentication in the near future.

The main application loop, as well as all the other 'front panel visible' functions, has routines to listen for commands received from the controlling client. When a command is received it is verified as a legal command, logged and then processed. Any changes caused by this command are broadcast to all clients. The server also has routines to detect changes to variables or a local operator command/selection. When this is encountered the status and/or changes of the server application are broadcast to all connected clients.

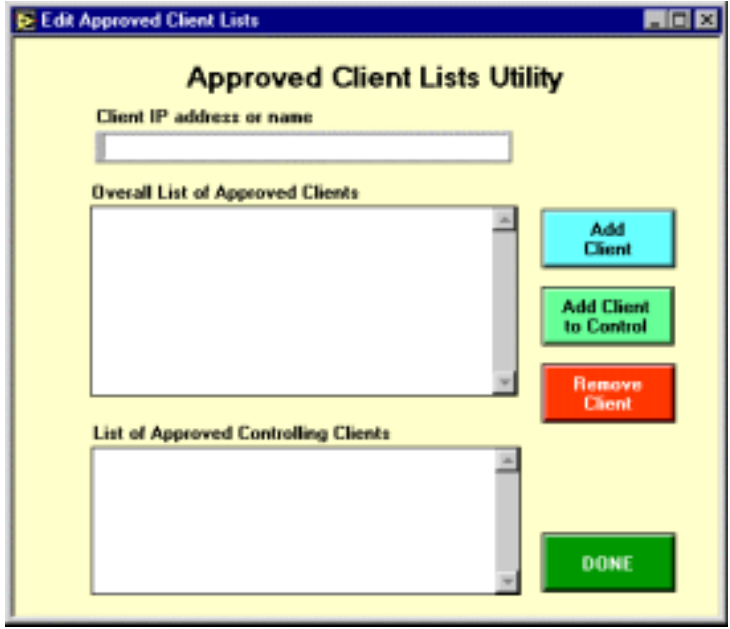

Figure 3 Client list utility front panel.

**Client Application**

The Client application operates in one of two modes: observer or control mode.
- In the **Observer mode**, the client application can only receive data from the server. The Client does not have control over the server. Commands sent to the server are ignored; except for the message indicating the client application is terminating and therefore closing the connection.
- In the **Control mode** the client application, besides receiving data from the server, can send commands to the server and essentially control the operation of the Server Application. In order to be in Control mode, the Client, after establishing the connection, must request control privileges and must be approved by the Server. Only one Client can be a controlling client at any one time. The server can refuse to grant control, and can also temporarily or permanently terminate control privileges.

The client relies on its own routines to display the user interface of the application as a reflection of the server. When a command is sent, the client operation normally waits for "change of application status" communication from the server before activating the change in its display.

**Functions and Network Commands**

In the current implementation, there is no acknowledgment of communication reception sent to the sender. The clients know that the server has received a command by the change in application state or change in parameter value. The server has no knowledge that the clients have received its broadcast. All command packets consist of a 4-byte number representing the length (in bytes) of the message followed by the message. The message is in ASCII.

**Command/Response Structure**

The command/response follow a structure similar to that used in GPIB instruments. Command lists are stored in global arrays in both applications. There are two main categories of commands: Active Function Commands and Global Commands. For Active Function Commands the first field identifies the **section** of the program (active function) the message is targeted to, so that if the server application is in another state, it can ignore that message. Any active section of the server recognizes the Global Commands.

Each command consists of a combination of fields separated by a (:). There are no spaces in the message, except within string values. Multiple parameters can be included in the same message. They consist of the entire command/response followed by a new line character. The commands are not case sensitive, i.e. they can be in upper case, lower case or a combination of both. The application automatically converts all messages to upper case.

| Structure of Active Function Commands | |
|---|---|
| Simple command message | `Section:command` |
| Simple variables | `Section:parameter=value` |
| 1-Dl Array (F=full, S=single) | `Section:F:1D-array=value,value,value,`<br>`Section:S:1D-array=index,value` |
| 2D arrays are sent as multiple message of 1D arrays. | `Section:F:2D-array=row-index,value,value,value,`<br>`Section:S:2D-array=row-index,column-index,value` |
| Structures | `Section:Structure:variable=value;variable=value;` |
| 1D Array of structures | `Section: 1D Array`<br>`Structure:index;variable=value;variable=value;` |

| Global Commands | |
|---|---|
| `REMQUIT` | This command indicates to the Server that a client has terminated and therefore its connection needs to be closed (at both ends). Any active section of the Server can perform this function and remove the associated client from the active connection lists. |
| `LOCQUIT` | This command indicates to the clients that the Server has terminated and therefore its connection needs to be closed (at both ends). |
| `STATUS` | This command is needed by the Clients to synchronize their data and display with that of the server. A Client can start at any time, and therefore needs this information. Status information consists of the values of all the globals, one of which is the active function, and the list of functions and/or globals that have their front panel visible on the server screen. |
| `GVAR` | This is the prefix for a change in global variable command. The Server uses it to send status information or to broadcast the change of a global variable. The control Client uses it to send the change to the Server. When the Server receives this change, it must rebroadcast the change to all other clients. The Client sending the change command does not implement its change until it receives notification from the server. |

| Structure of Global Commands | |
|---|---|
| Simple global commands | `Command` |
| Global variable value change command. Normally, GVAR is followed by a global group code indicating the VI that has multiple variables. | |
| Simple variables | `GVAR:gg:parameter=value` |
| 1-D Array (F=full, S=single) | `GVAR:gg:F:1D-array=value,value,value,`<br>`GVAR:gg:S:1D-array=index,value` |
| 2D arrays are sent as multiple commands of 1D arrays. | `GVAR:gg:F:2D-array=row-index,value,value,value,`<br>`GVAR:gg:S:2D-array=row-index,column-index,value` |
| Structures | `GVAR:gg:Structure:variable=value;variable=value;` |
| 1D Array of structures | `GVAR:gg:1DArrayStructure:index;variable=value;variable=value;` |

**"Run the System" Main Program**

This version of the program assumes that the Server has started and is in the main loop waiting for a command. There are a couple of initial steps in which the program is waiting for user input. These steps are not currently controlled remotely (they could be). It is assumed that a local user initializes the Server.

When the Server application is started, it performs some initial steps and then waits for a user command, either local or from a remote connection. The main program listens to the commands from the controlling computer. When started, the Client tries to establish a connection with the Server. If it is not successful, whether because the Server is not running or the network is down, the client terminates. If successful it can operate in one of two modes: Controller or Observer. In Observer mode the client initially requests a status to synchronize its data and display with the server. Then, it can only receive data from the Server. Commands sent are ignored by the server, except for the REMQUIT command indicating that this client is terminating. In the Controller mode the client initially requests a status to synchronize its data and display with the server. Then, it can observe or send commands and/or change variables, etc. to the server. When the selected function in the Server becomes active, it will broadcast to all active network connections that it is active and the values of the variables on its display.

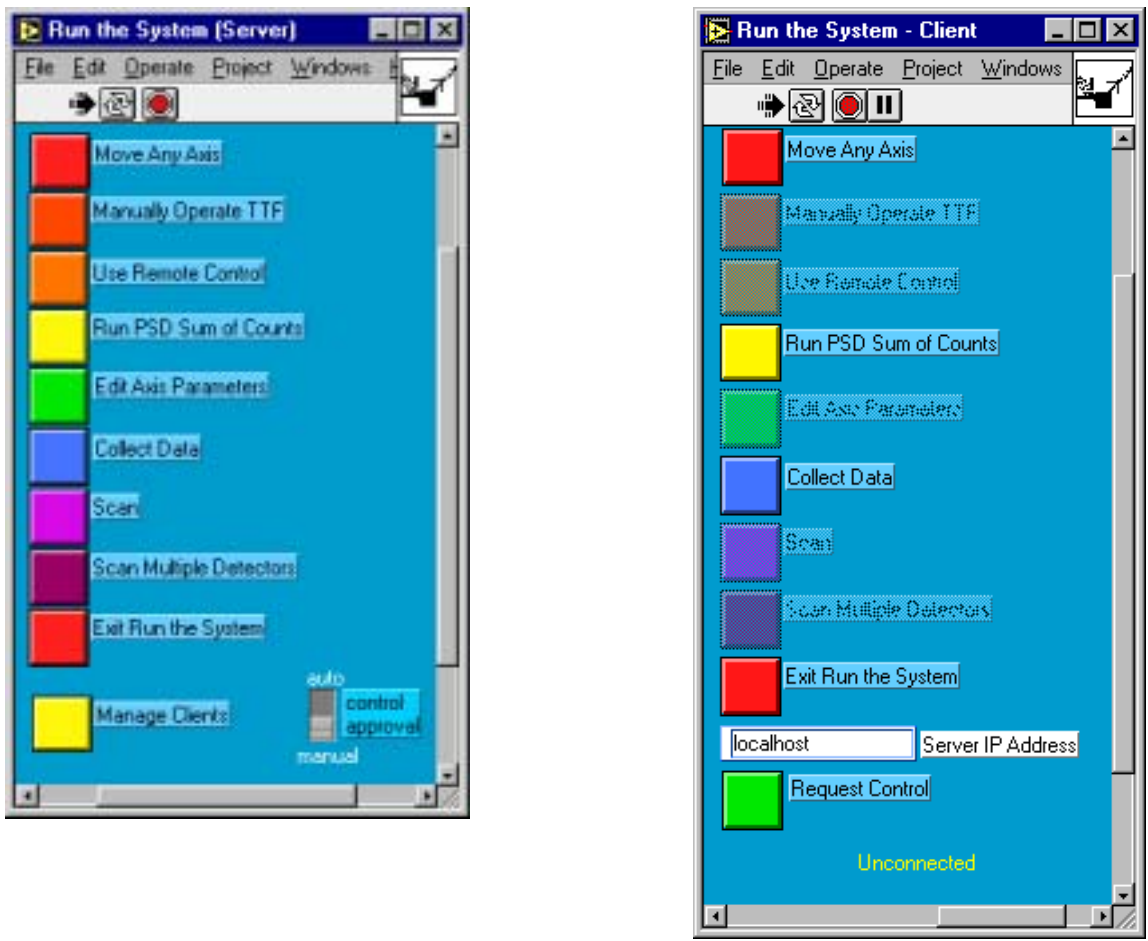

Figure 4 Top-level front panels for the server and client applications. The appearance of the front panels of the second level functions is identical on the client and the server.

## Video Everywhere

We have made extensive use of streaming JPEG video in a web browser. For most applications, we have found this technique, combined with the telephone, to be more useful for daily interaction than traditional video conferencing over the Internet. These video streams perform remarkably well and require no special software on the receiving end. Using frames, multiple video images can be display on a single web page. We have recently begun switching from PC's with frame grabber cards to Axis[8] network cameras and video servers. These devices have built-in web servers on a chip and connect directly to the network without the need for a frame grabber and PC.

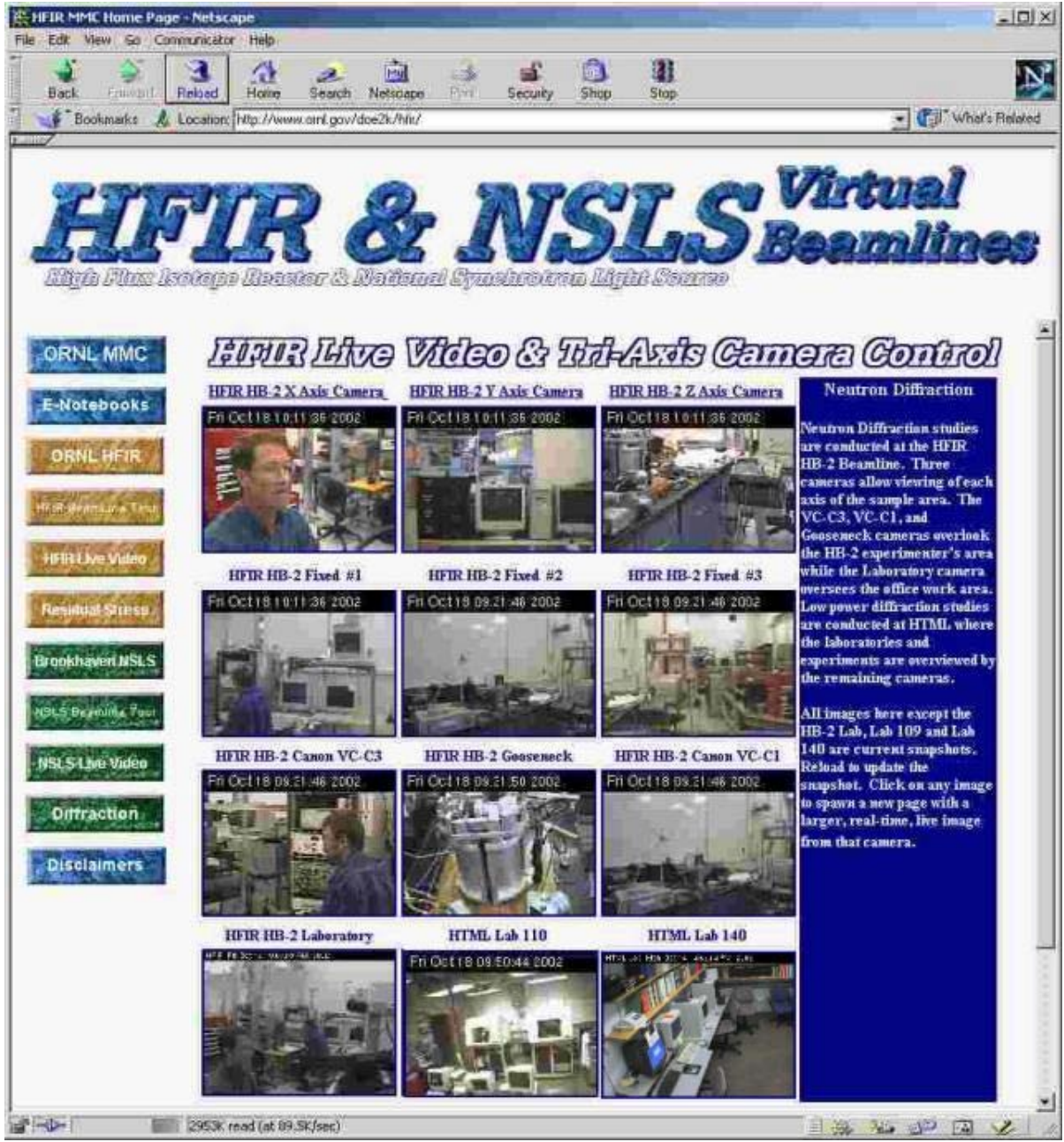

Figure 5 An example of using streaming video to create a sense of "being there".

Another example of the usefulness of streaming, Figure 5, shows how these images can be used to create a sense of "being there". The page shows static images of several labs and common areas. Any or all of the images can be converted into live video by simply clicking on the image. This is the online equivalent of walking down the hall and entering an office or a lab to see what is happening there. In the MMC project we have about 50 video feeds available at all times in labs and offices around the country. The desire is to simulate being "down the hall while around the world".

## Electronic Notebooks

If scientists are to be able to work together they must be able to share access to the experimental logbooks. One of the DOE2000 technology R&D projects is centered around the development of an Electronic Notebook[9]. The notebook is a collection of web pages that are created and viewed through a browser. Electronic notebooks can be:
- Shared by a group of researchers,
- Accessed remotely,
- Immune from being misplaced, lost, or accidentally destroyed (if backed up),
- Data repositories containing computer files, plots, etc,
- Easily searched for information,
- Multimedia rich — audio/video clips,
- Linked to other information.

We have been using the electronic notebook for both local and remotely operated experiments at NRSF. Figure 6 shows a sample page from the notebook. Input to the notebook can come from:
- Text entered into a dialog box,
- Files stored on disk,
- Graphics captured from the computer screen,
- Sketches drawn on a graphics tablet,
- Operating parameters electronically acquired directly from the instrument.

New notebooks can be created online by authorized users, and they then control access lists for their notebook.

The notebook system is interfaced to the user via a set of Java servlets that sit between the web server and the user's browser. Every URL request is automatically secured with no change required to the web server. The servlets grab the user's PKI certificate and uses that to provide strong authentication and also access control  However, all of this is transparent to the user, and there is actually less hassle than with a user id/password-protected Web server. The Servlet-based notebook system has proven to be extremely reliable. In over a year of use there has not been a single problem reported by the users.

The notebook is written in Perl. About 10 lines of Perl have to be modified to separate the naming and location of the notebook from the code that actually does the work. In this way, new notebooks can be created and accessed via the servlet front end. Version 2 of the notebook is written in Java. We hope to switch to a fully Java-based version in the coming year.

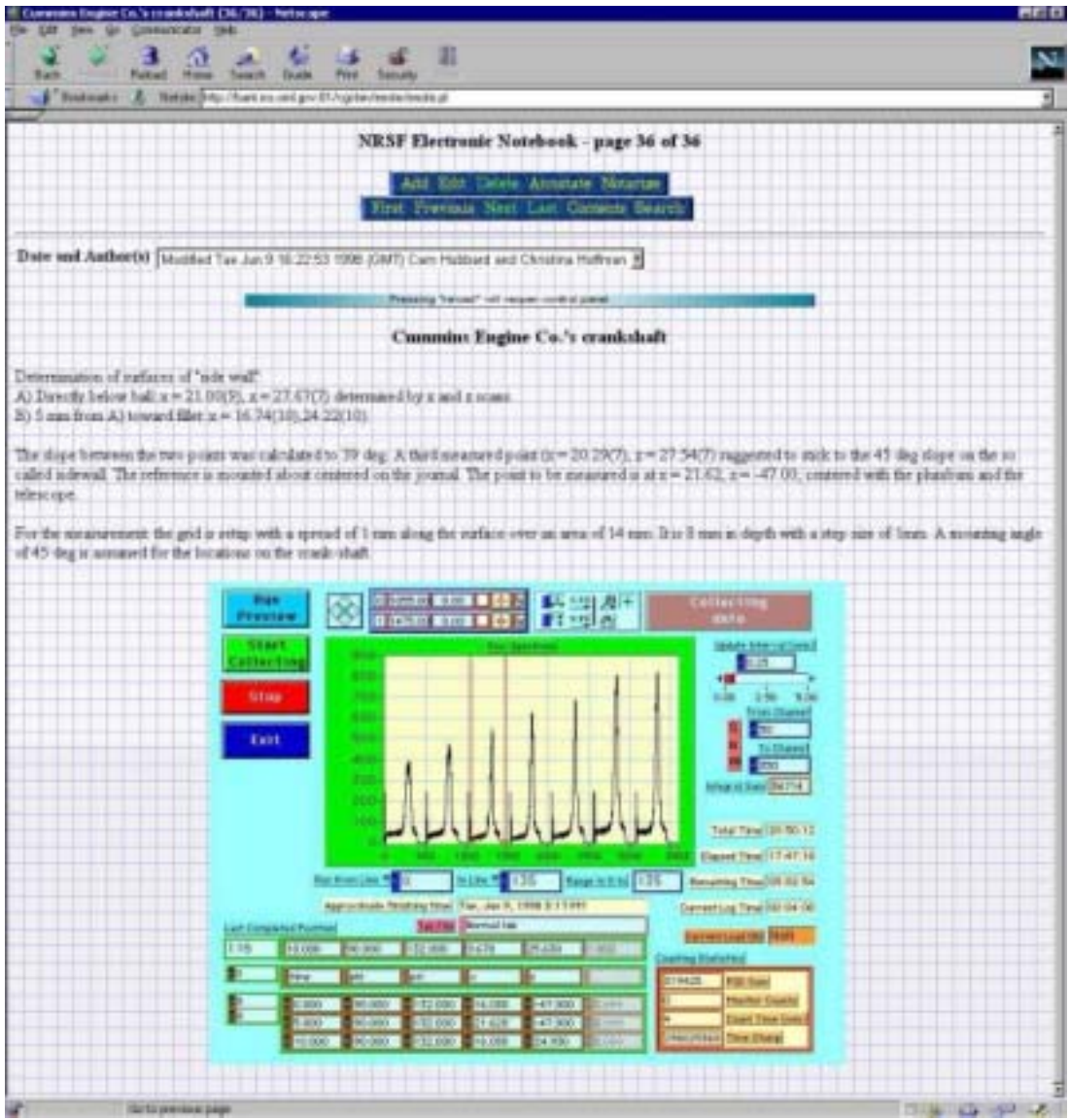

Figure 6 A sample page from the electronic notebook in use at NRSF.

## Security

Security is vital on the Internet. We are putting valuable and complicated facilities online to the whole world. If security "gets in the way" it will not be used or will inhibit collaboratory goals but if it is too weak, valuable assets will be at risk. A full discussion of our security plans is beyond the scope of this paper but further discussion can be found through our project web site. Briefly, we have been developing and beginning to deploy an X.509 certificate system to control access to our facilities and to encrypt the information traversing the network. Certificates are now supported by modern Web browsers but are not yet widely used. However, the need for strong security for electronic commerce on the web is driving major developments in this field and the certificate infrastructure and tools have greatly improved this past year. We are using authority certificates in addition to authentication certificates to provide more fine-grained control over the functions that a remote user is allowed to perform on the instrument.

A growing challenge in the post 9/11 world is to reconcile the security needs of our collaboratory with those of both our organizations and those of our users. Firewalls often block remote control programs (Timbuktu and VNC) and local policies often disallow encrypted internal traffic. Efforts are underway to create security enclaves that can cross organizational boundaries, while at the same time satisfying all security policies.

## Summary and Next Steps

The three approaches to remote control (Web-based, remote computer control, and client-server) are all useful for control of different instruments and even for different control requirements on a particular instrument. We are using all three methods for the Neutron Residual Stress Facility. While the most difficult to implement, the client-server approach provides the highest level of performance and security.

We also utilize remote scientific collaboration and not just remote instrument control. Secure electronic notebooks are an important tool for doing that. The quality of live video over the Internet is improving and it has become inexpensive and easy to install. Commercial software like LabVIEW, and even Microsoft Office, are increasingly network-aware making remote collaboration easier. We have been doing Internet-based remote control since 1994 and the changes in the availability and performance of software tools and standards during this time have been enormous. Several things we formerly had to do "by hand" are now available commercially. When the new instruments are operational we will revisit some of the decisions we have made earlier in light of the availability of these new tools and standards.

## Acknowledgements

This work was begun as part of the Materials Microcharacterization Collaboratory (MMC), a pilot project within the US Department of Energy's DOE2000 program. Subsequent research sponsored by the Assistant Secretary for Energy Efficiency and Renewable Energy, Office of Transportation Technologies, as part of the High Temperature Materials Laboratory User Program, Oak Ridge National Laboratory, managed by UT-Battelle, LLC, for the U.S. Dept. of Energy under contract DE-AC05-00OR22725. This paper mentions commercial products that we have used, but should not be considered an endorsement of these products.